\documentclass[cits]{PoS}
\usepackage{subfigure}
\usepackage[utf8]{inputenc}
\usepackage{units}
\usepackage{amsmath}
\usepackage{textcomp}
\usepackage{amsfonts,amssymb}
\usepackage{caption}
\title{Semileptonic $B$ / $B_s$ decays at Belle}

\ShortTitle{Semileptonic $B$/$B_s$ decays at Belle}

\author{\speaker{Christian Oswald} 
	\\
        University of Bonn\\
        E-mail: \email{oswald@physik.uni-bonn.de}}

\abstract{The Belle experiment at the KEKB asymmetric energy $e^+e^-$ collider recorded  large data sets of both, $B$ and $B_s$ decays. Semileptonic decays $B_{(s)} \to X \ell \nu$ ($\ell=e$ or $\mu$) constitute approximately one fifth of the total decay width of $B_{(s)}$ mesons and play an important role in the determination of the CKM matrix elements $|V_{ub}|$ and $|V_{cb}|$. Recent results from Belle are presented, including the study of $B^- \to D_s^{(*)} K \ell \nu$, the first measurements of semi-inclusive modes $B \to D^{(*)} X \ell \nu$ and the measurement of the inclusive branching fraction $\mathcal{B}(B_s \to X \ell\nu)$.}

\FullConference{36th International Conference on High Energy Physics\\ 4-11 July 2012\\ Melbourne, Australia}

\begin{document}

\section{Introduction}
Semileptonic decays of $b$-flavoured mesons ($B \to X \ell \nu$) play a crucial role in the determination of the CKM matrix elements $|V_{ub}|$ and $|V_{cb}|$. Measurements are performed either by summing over all possible final states (inclusive approach) or by explicitly reconstructing an hadronic final state, e.g. a $D$ meson (exclusive approach). Both approaches are complementary, relying on different experimental and theoretical techniques.

Despite the excellent precision of the measurements of the inclusive semileptonic $B^+$ and $B^0$ widths\footnote{Charge conjugation is always implied if not explicitly stated otherwise.}, not all features are yet completely understood. There is a discrepancy between the sum of the exclusive rates $B \to D^{(*,**)} \ell\nu$  and the inclusive rate $B \to X_c \ell\nu$, called the ``inclusive vs. exclusive puzzle''. One candidate for this gap could be $B^- \to D_s^{(*)} K \ell \nu$, however, the branching fractions are too small. A recent theory paper \cite{Bernlochner:2012bc} proposes that the missing part of the semileptonic width consists of radially excited $D^{\prime (*)}$ mesons. A new set of semi-inclusive measurements $B \to D^{(*)} X \ell \nu$ was performed at Belle to provide new experimental insight.

In contrast to $B^+$ and $B^0$ mesons, the knowledge of the semileptonic decays of $B_s^0$ mesons is rather sparse. Some exclusive modes, $D_{s1}(2536)X$ and $D_{s2}(2573)X$, were measured recently by the D0 and LHCb collaborations \cite{Abazov:2007wg, Aaij:2011ju}. The BaBar collaboration published a measurement of $\mathcal{B}(B_s^0 \to X \ell \nu)$ \cite{babarbsxlnu2012}. The inclusive semileptonic branching fraction  is an important normalisation mode for other measurements \cite{Aaij:2011jp}. A widely used expectation from SU(3) symmetry is that the total semileptonic widths of the $B_{(s)}$ mesons are equal: $\Gamma_\text{sl}(B_u^+) = \Gamma_\text{sl}(B_d^0) = \Gamma_\text{sl}(B_s^0)$. Theory calculations predict indeed that higher order terms due to QCD effects vanish and the expected deviation is only at the percent level \cite{bigi2011, Gronau2010}. 

\section{The exclusive decay $B \to D_s^{(*)} K \ell \nu$ \cite{Stypula:2012mf}}
The measurement of $B \to D_s^{(*)} K \ell \nu$ is based on a sample containing approximately $657 \times 10^6$ $B\bar{B}$ pairs collected at the $\Upsilon(4S)$ resonance. Signal candidates are formed by
combining a $K^-$ and $\ell^-$ candidate with a $D_s$ meson reconstructed in the mode $D_s \to \phi \pi^+,~\phi \to K^+ K^-$. $D_s^{*}$ mesons are reconstructed via $D_s^{*} \to D_s \gamma$, with the photon energy $E_\gamma > \unit[125]{MeV}$. The existence of the invisible neutrino is inferred from  the known beam energy $E_\text{beam}$, the $B$ meson mass $M_B$ and the visible four-momentum $(E_\text{vis}, \vec{p}_\text{vis})$ of the reconstructed signal particles. 
\begin{figure}
\centering
\subfigure[$|X_\text{mis}| < 1$]{
\includegraphics[width=0.2\textwidth]{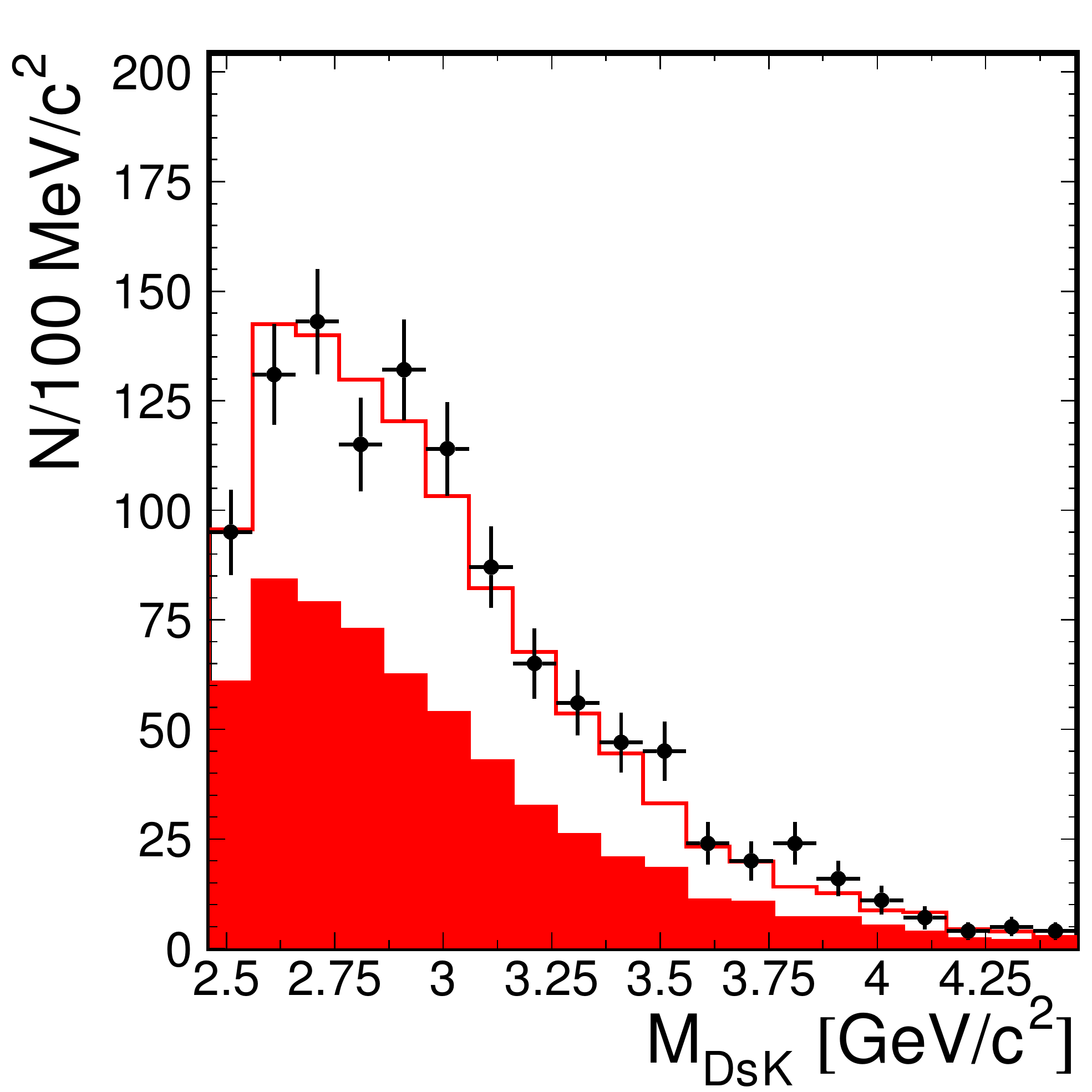}
\label{fig:mdsk_bg}
}
\subfigure[$|X_\text{mis}| > 1$]{
\includegraphics[width=0.2\textwidth]{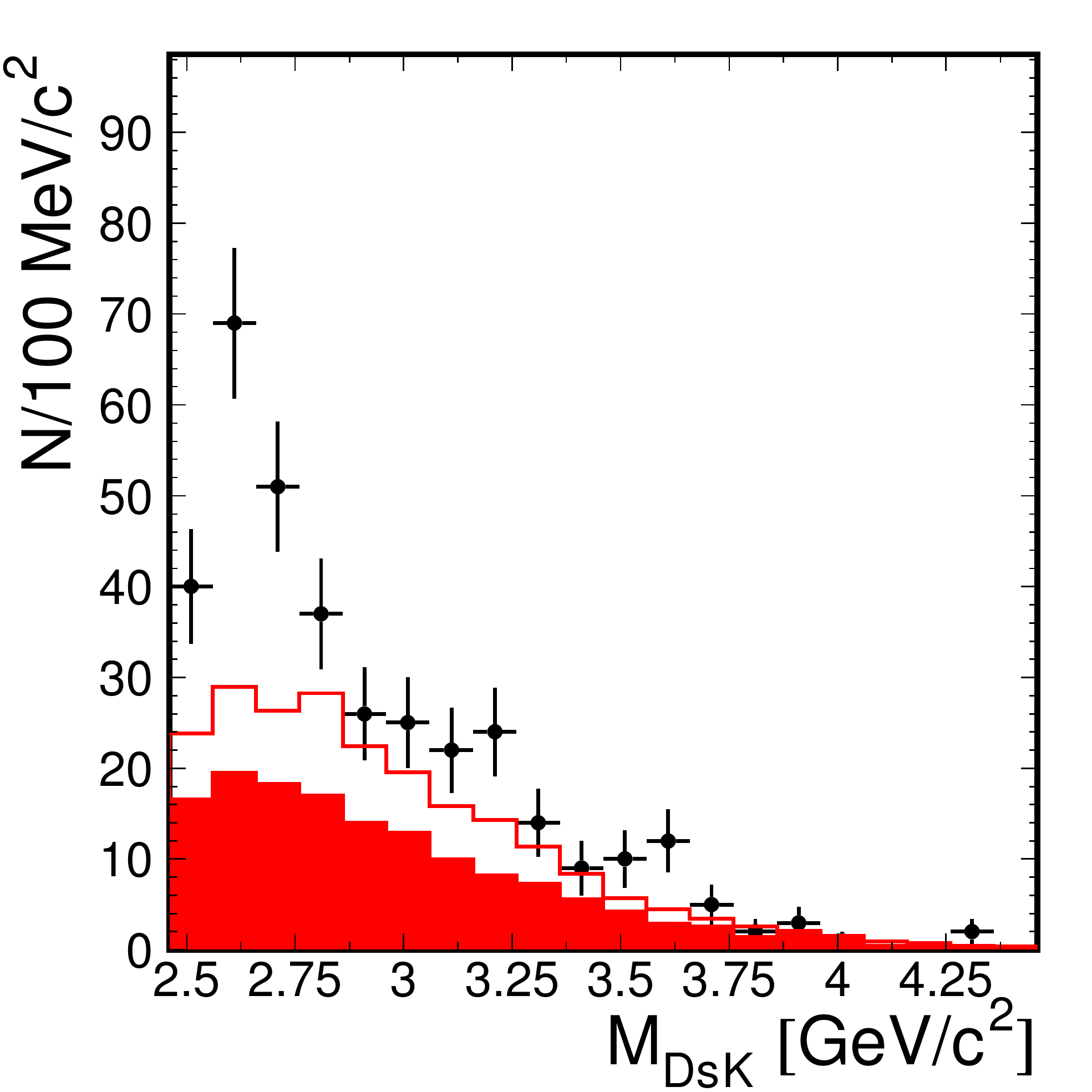}
\label{fig:mdsk_sig}
}
\caption{$D_sK$ mass distributions. The black points are the data and the histograms show the expected background contributions from fake $D_s$ (full) and true $D_s$ (blank). The histograms are superimposed additively. \cite{Stypula:2012mf}}
\label{fig:mdsk}
\end{figure} 
The variable $X_\text{mis} = \left[(E_\text{beam} - E_\text{vis}) - |\vec{p}_\text{vis}\right] / \sqrt{E^2_\text{beam} - M^2_B}$ lies for correctly reconstructed semileptonic decays in the interval $[-1,1]$. Figure \ref{fig:mdsk} shows the $D_sK$ mass distributions in the signal region $|X_\text{mis}|<1$, and the side band. The most problematic background in this analysis consists of events where the reconstructed signal contains particles from both $B$ mesons ($B^- \to \ell^- X$ and $B^+ \to D_s^+ X^\prime$). This background is suppressed by a semileptonic tag where an additional high momentum lepton, $p(\ell^+) > \unit[0.5]{GeV}$, of the opposite charge as the signal lepton $\ell^-$ is reconstructed. Tracks and clusters in the electromagnetic calorimeter not used for signal reconstruction are exploited to determine variables for further background rejection: the hadron mass on tag side $M^c_\text{tag}$, which is expected to be around the $D^{(*)}$ mass, and the variable $X_\text{tag}$, which is the 
counterpart of $X_{mis}$ on the tag side.

\begin{figure}
\subfigure[]{
\includegraphics[width=0.17\textwidth]{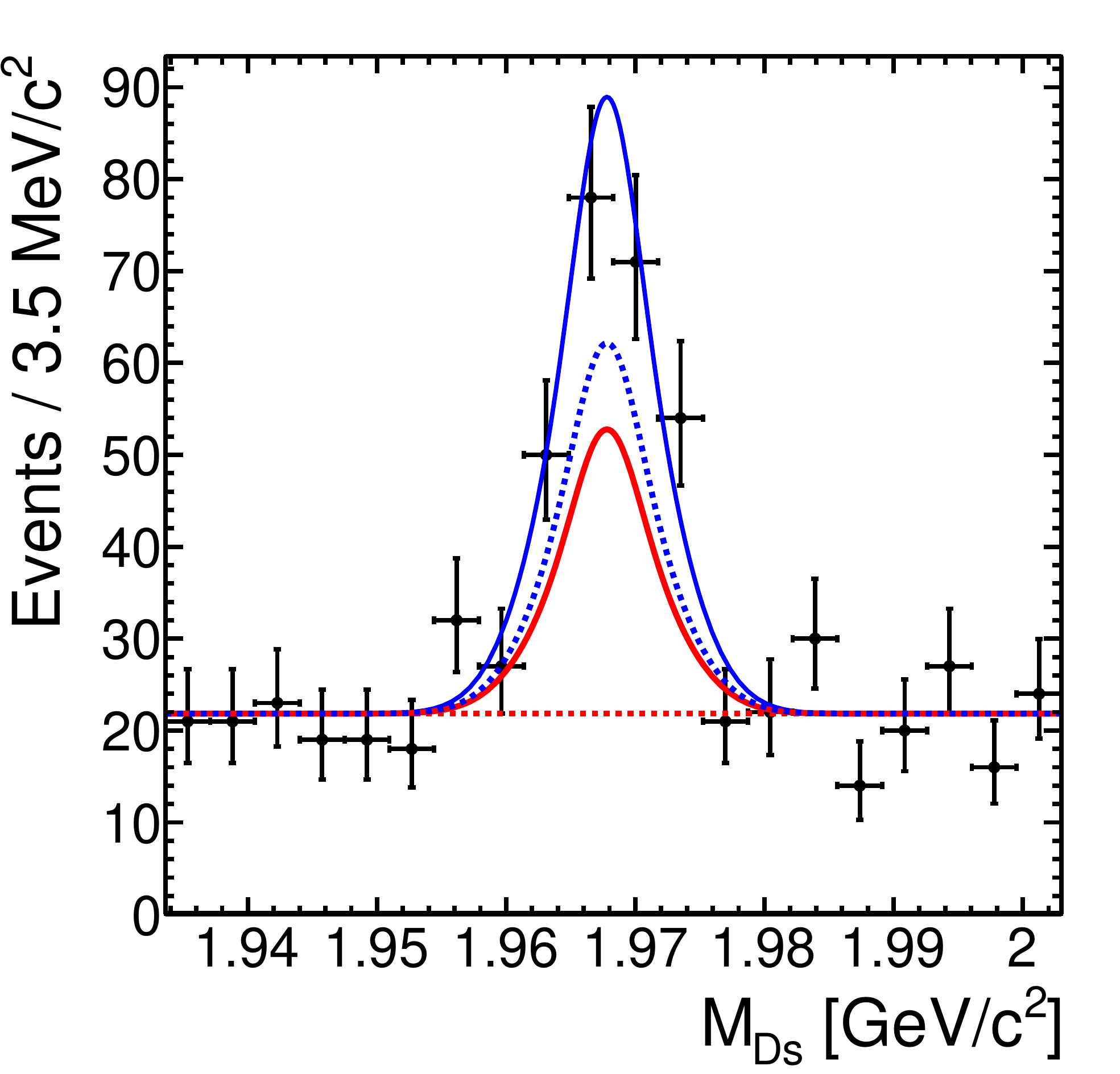}
\label{fig:fitX_mds0}
}
\subfigure[]{
\includegraphics[width=0.17\textwidth]{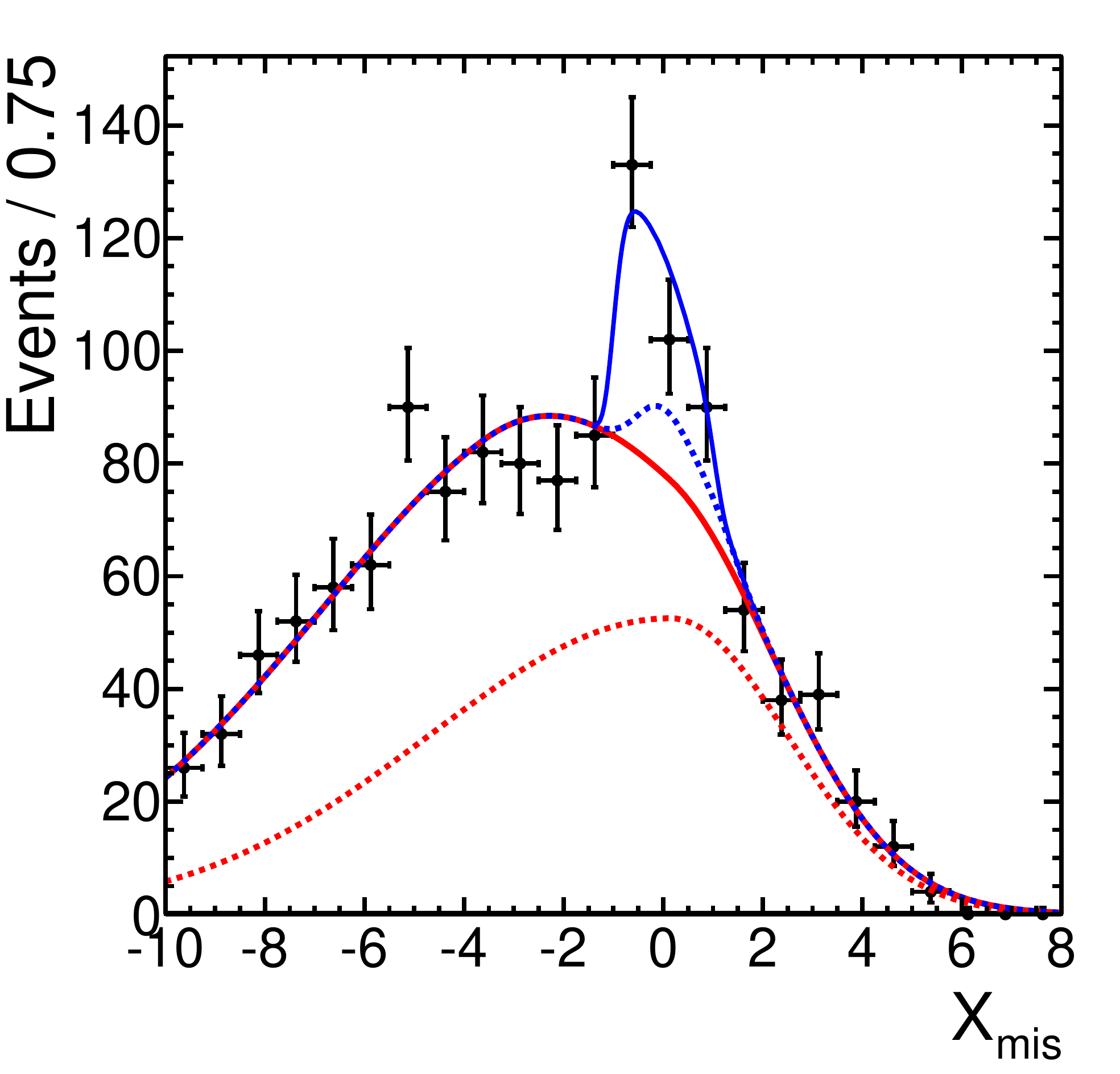}
\label{fig:fitX_xmis0}
}
\subfigure[]{
\includegraphics[width=0.17\textwidth]{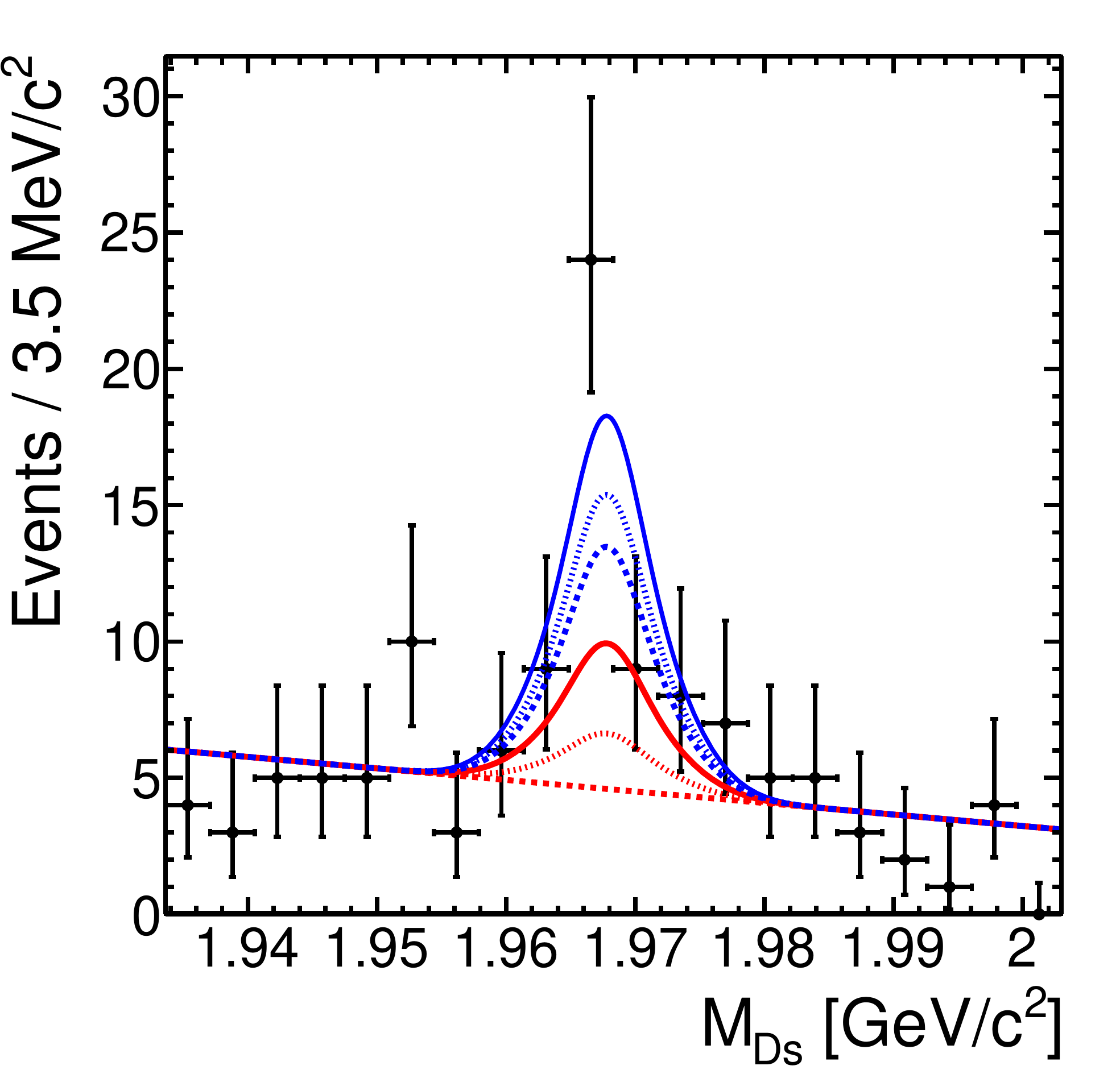}
\label{fig:fitX_mds1}
}
\subfigure[]{
\includegraphics[width=0.17\textwidth]{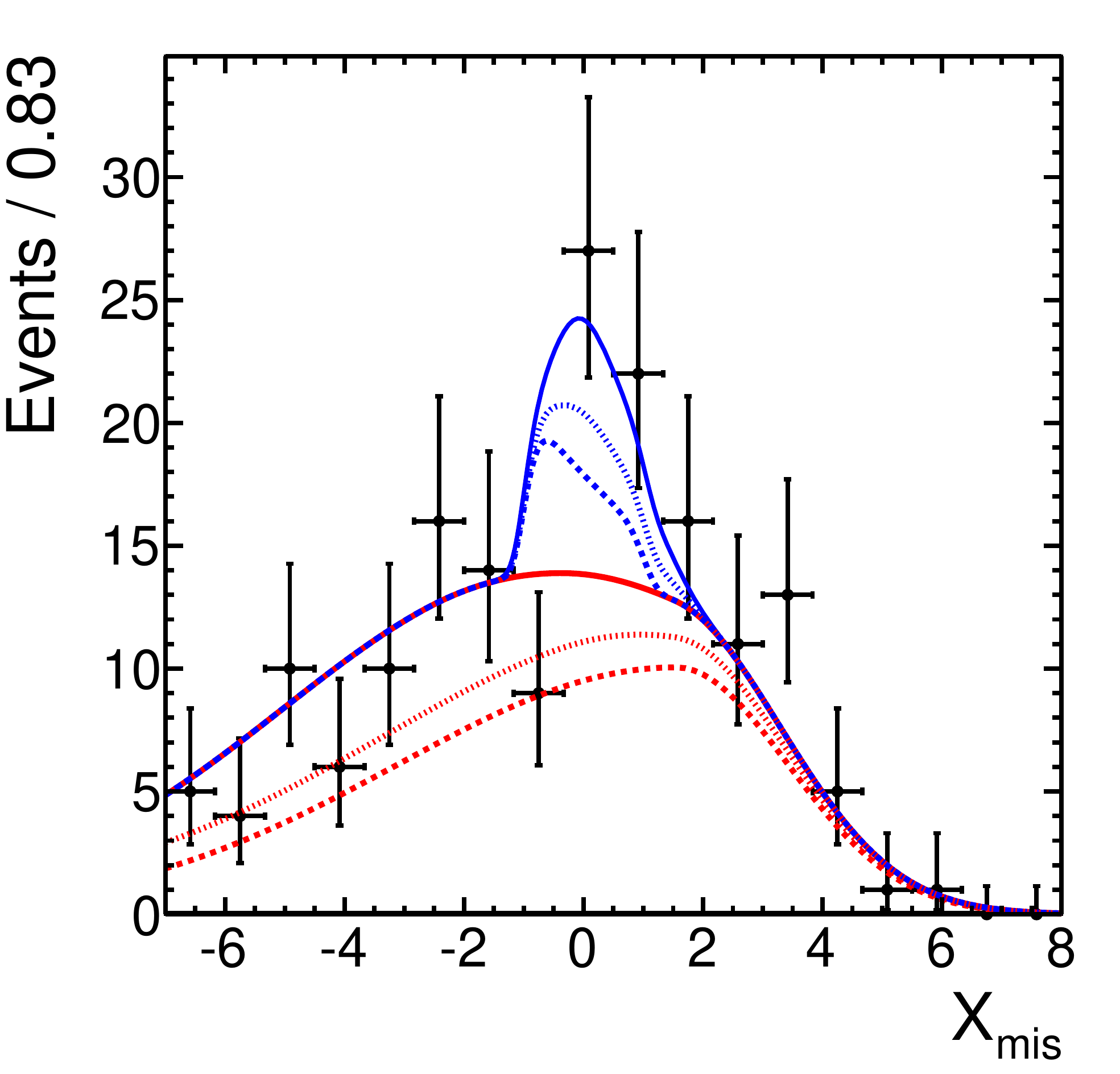}
\label{fig:fitX_xmis1}
}
\subfigure[]{
\includegraphics[width=0.17\textwidth]{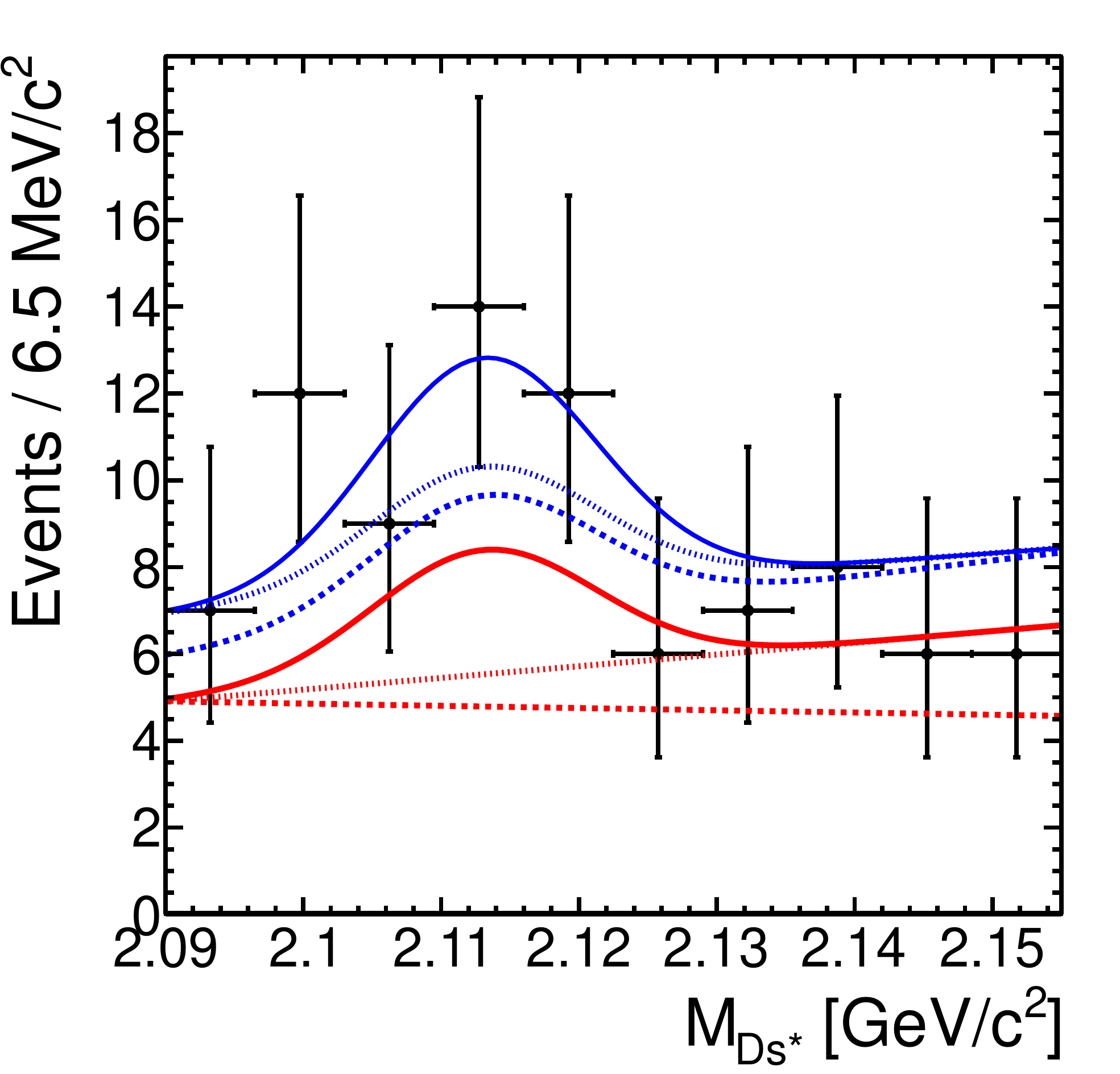}
\label{fig:fitX_mdss}
}
\caption{Fit projections (lines) and data (black points). The variables are shown in the signal window of the other variable(s). The fit components of the $D_s^+$ sample (a, b) are from bottom to top: fake and true $D_s^+$ from background, signal contributions from $D_s^{*+}$ and $D_s^+$. The fit components of the $D_s^{*+}$ sample (c, d, e) are from bottom to top: background contributions from fake $D_s^+$, fake $D_s^{*+}$ and true $D_s^{*+}$, and signal contributions from the $D_s^+$ mode, the $D_s^{*+}$ mode with fake $D_s^{*+}$ and with true $D_s^{*+}$. The fitted contributions are superimposed additively. \cite{Stypula:2012mf}}
\label{fig:fitX}
\end{figure}

The branching fractions are extracted in a five dimensional simultaneous unbinned maximum likelihood fit (see Fig. \ref{fig:fitX}) to the variables $X_\text{mis}$ and $m(D_s)$ for the $D_sK$ mode, and $X_\text{mis}$, $m(D_s)$ and $m(D_s^*)$ for the $D_s^{*}K$ mode. Cross feed between the modes is taken into account. The obtained branching  fractions are $\mathcal{B}(B^- \to D_s K \ell \nu) = (0.30 \pm 0.90_\text{~stat.} \pm^{0.11}_{0.08~\text{sys}}) \times 10^{-3}$ with an significance of $3.9 \sigma$, and $\mathcal{B}(B^- \to D_s^{*} K\ell \nu) < 0.56 \times 10^{-3}$ at the $90\%$ confidence level. This is the first measurement of the separate $D_s K\ell\nu$ and $D_s^{*}K\ell\nu$ modes. The combined mode $D_s^{(*)}K\ell\nu$ is observed with a significance of $6 \sigma$ and its branching ratio is $\mathcal{B}(B^- \to D_s^{(*)} K\ell\nu) = (0.59 \pm 0.90_\text{stat.} \pm^{0.11}_{0.08, \text{sys}}) \times 10^{-3}$.

\section{Study of semi-inclusive semileptonic $B^+$ and $B^0$ decays}
\begin{figure}
\centering
\subfigure[]{
\includegraphics[height=0.2\textwidth]{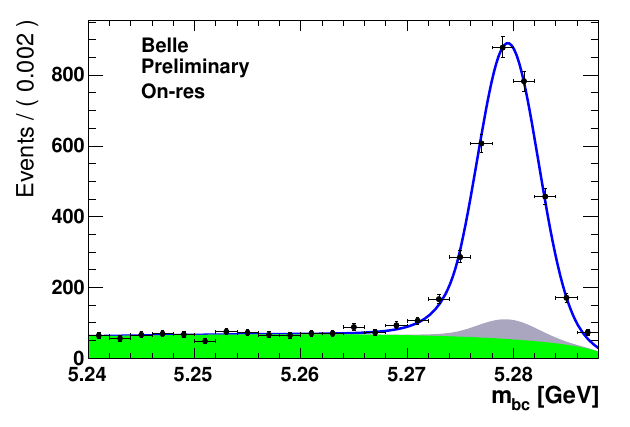}
\label{fig:semiinc_mbc}
}
\subfigure[]{
\includegraphics[height=0.2\textwidth]{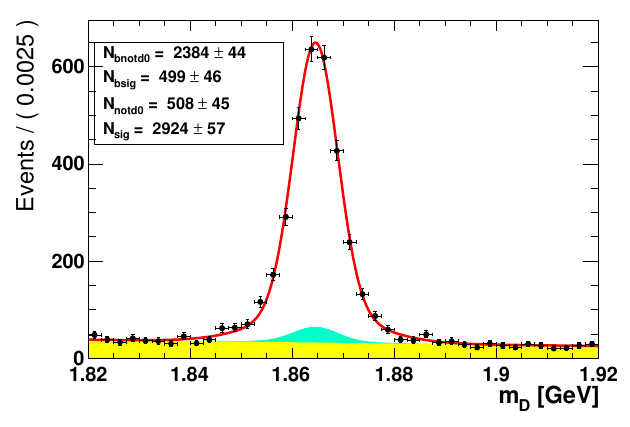}
\label{fig:semiinc_md}
}
\subfigure[]{
\includegraphics[height=0.2\textwidth]{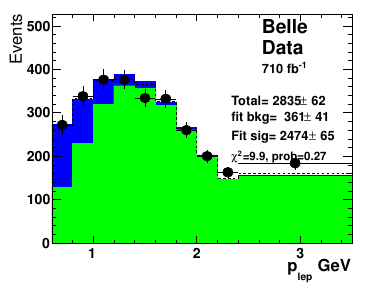}
\label{fig:semiinc_plep}
}
\caption{Fits of the semi-inclusive analysis at the example of the $\bar{B}^0 \to D^0 \ell \nu X$ mode. Black points with error bars are data. \emph{(a):} Beam constraint mass with misreconstructed tag (green), misidentified $D^0$ (grey) and signal (blue). \emph{(b):} Reconstructed $D^0$ mass with misidentified $D^0$ (yellow), isreconstructed tag (teal) and signal (red). \emph{(c)} Lepton momentum spectrum with fitted prompt lepton component (green), secondary and misidentified leptons (blue).}
\label{fig:semiinc}
\end{figure}

\begin{figure}
\centering 
\begin{minipage}{0.45\textwidth}
\footnotesize
\begin{tabular}{lll}
\hline \hline
$B$ & $X^\prime$   & $\mathcal{B}(B \to X^\prime \ell \nu) / \mathcal{B}(B \to X \ell \nu)$ \\
\hline 
$B^-$&$D^0 X$ & $0.922 \pm 0.016_{\rm stat.} \pm 0.011_{\rm {\cal B}(D)} \pm 0.036_{\rm sys}$\\
$B^-$&$ D^+ X$ & $0.088 \pm 0.004_{\rm stat.} \pm 0.002_{\rm {\cal B}(D)} \pm 0.005_{\rm sys}$\\
$B^0$&$D^0 X$ & $0.575 \pm 0.016_{\rm stat.} \pm 0.007_{\rm {\cal B}(D)} \pm 0.022_{\rm sys}$\\
$B^0$&$D^+ X$ & $0.452 \pm 0.007_{\rm stat.} \pm 0.010_{\rm {\cal B}(D)} \pm 0.021_{\rm sys}$\\
$B^-$&$D^{*0} X$ & $0.597 \pm 0.026_{\rm stat.} \pm 0.007_{\rm {\cal B}(D)} \pm 0.024_{\rm sys}$\\
$B^-$&$D^{*+} X$ & $0.064 \pm 0.007_{\rm stat.} \pm 0.008_{\rm {\cal B}(D)} \pm 0.004_{\rm sys}$\\
$B^0$&$D^{*0} X$ & $0.081 \pm 0.020_{\rm stat.} \pm 0.009_{\rm {\cal B}(D)} \pm 0.006_{\rm sys}$\\
$B^0$&$D^{*+} X$ & $0.615 \pm 0.021_{\rm stat.} \pm 0.007_{\rm {\cal B}(D)} \pm 0.024_{\rm sys}$\\
\hline \hline  
\end{tabular} 
\captionof{table}{Preliminary results of the semi-inclusive analysis normalised to the total semileptonic width.}
\label{tab:semiincresults}
\end{minipage}
\hfill
\begin{minipage}{0.45\textwidth}
\includegraphics[width=\textwidth]{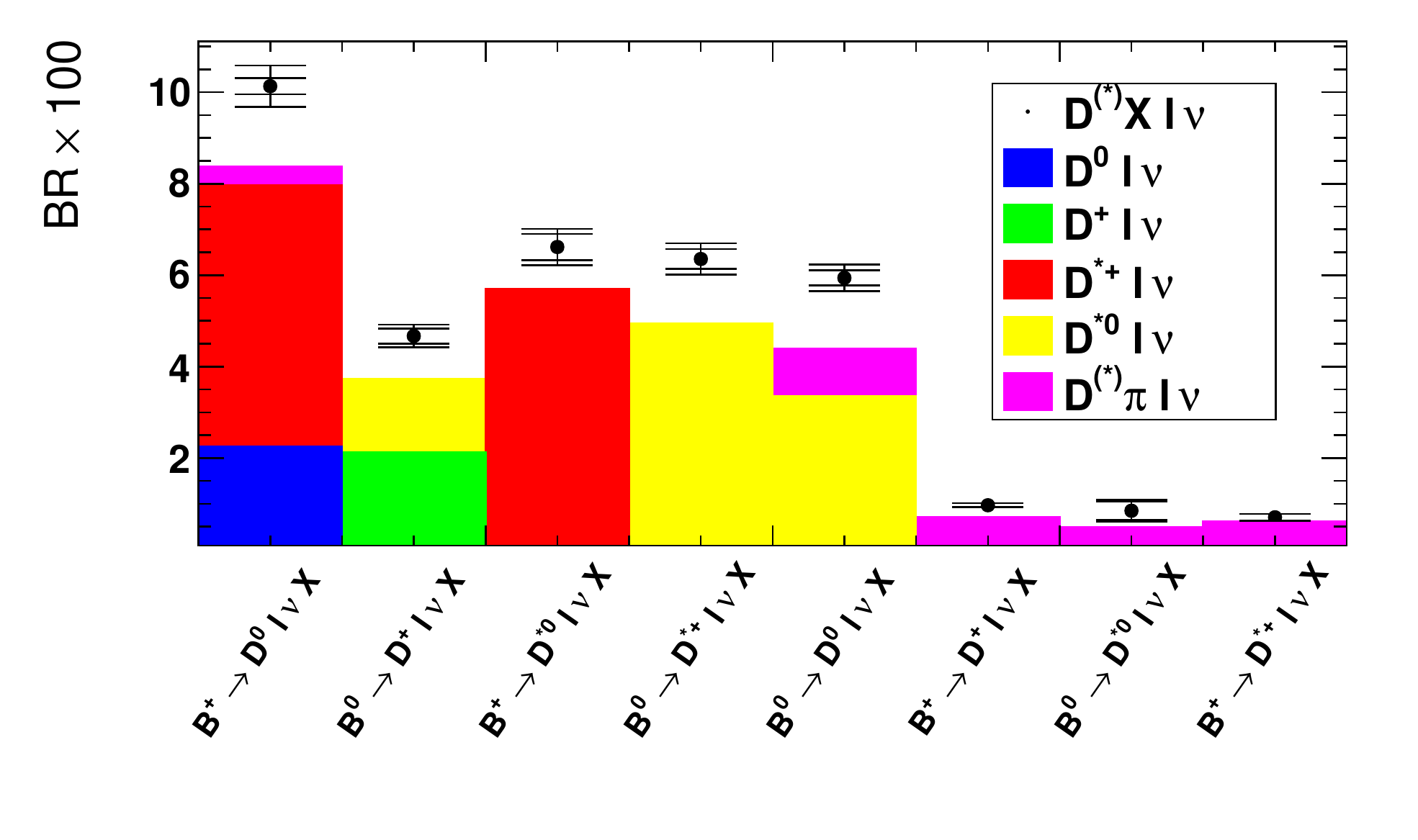} 
\caption{Semi-inclusive branching fractions measured at Belle (black points with error bars) compared to current world averages of exclusive measurements.} 
\label{fig:semiex}
\end{minipage}
\end{figure}

The semi-inclusive measurement is performed with the full ($\unit[710]{fb^{-1}}$) Belle data set collected at the $\Upsilon(4S)$ resonance. One $B$ in the event is fully reconstructed in hadronic modes with a neural network and serves as a tag to determine the four-momentum and flavour of the signal $B$. On the signal side, a lepton $\ell$ and a charmed meson ($D^0$, $D^-$, $D^{*0}$ or $D^{*+}$) are reconstructed. The number of signal events is extracted in a two dimensional fit to the beam energy constrained mass $m_\text{bc}$ and the mass of the reconstructed meson, $m_D$, in bins of lepton momentum; an example is shown in Fig. \ref{fig:semiinc} (a) + (b). The contamination from secondary and misidentified leptons is estimated from a $\chi^2$ fit to the lepton momentum spectrum (see Fig. \ref{fig:semiinc_plep}).
The results given in Tab. \ref{tab:semiincresults} are normalised to the inclusive semileptonic branching fraction $\mathcal{B}(B \to X \ell \nu)$ measured in the same $B$-tagged sample. This has the advantage that some of the systematic uncertainties cancel. The decay products of the unknown components of the semileptonic width can be inferred from the difference between the measured semi-inclusive branching fractions and the known exclusive modes (see Fig. \ref{fig:semiex}). 

\section{The inclusive semileptonic branching fraction of the $B_s^0$ meson \cite{bs2xlnu}}
\begin{figure}
\centering
\subfigure[]{
\includegraphics[width=0.3\textwidth]{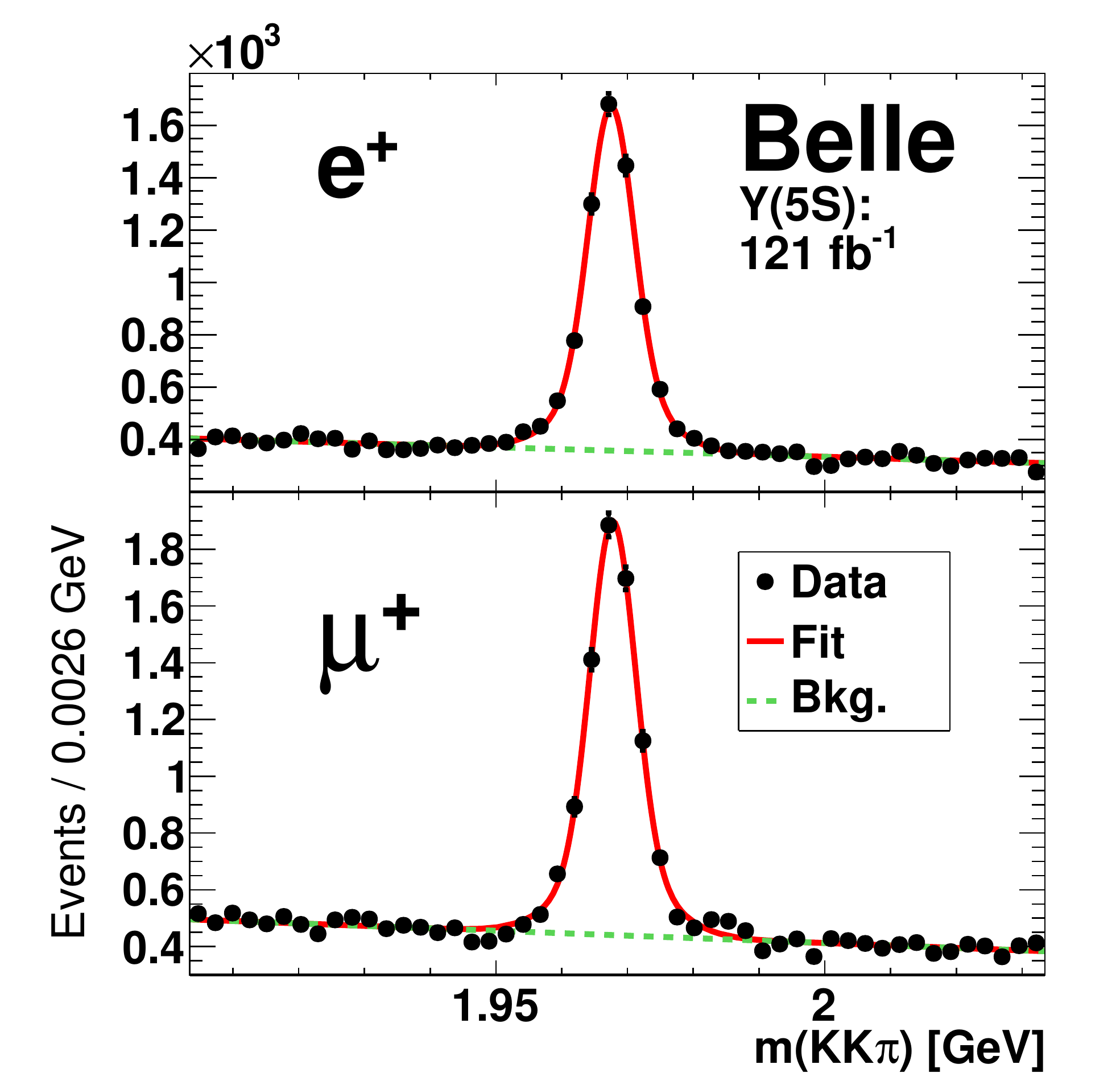}
\label{fig:kkpimassfit}
}
\subfigure[]{
\includegraphics[width=0.3\textwidth]{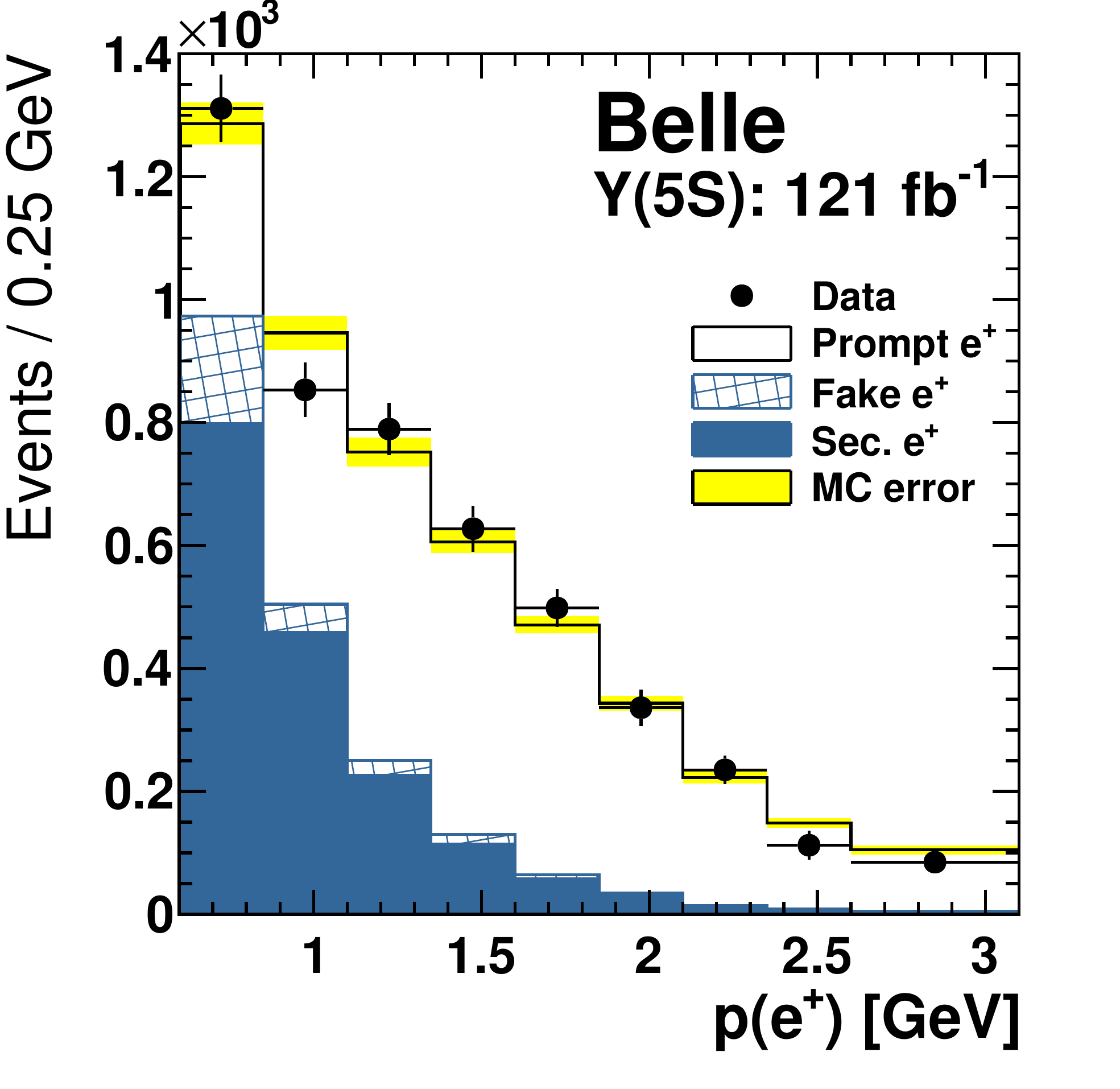}
\label{fig:elecmomfit}
}
\subfigure[]{
\includegraphics[width=0.3\textwidth]{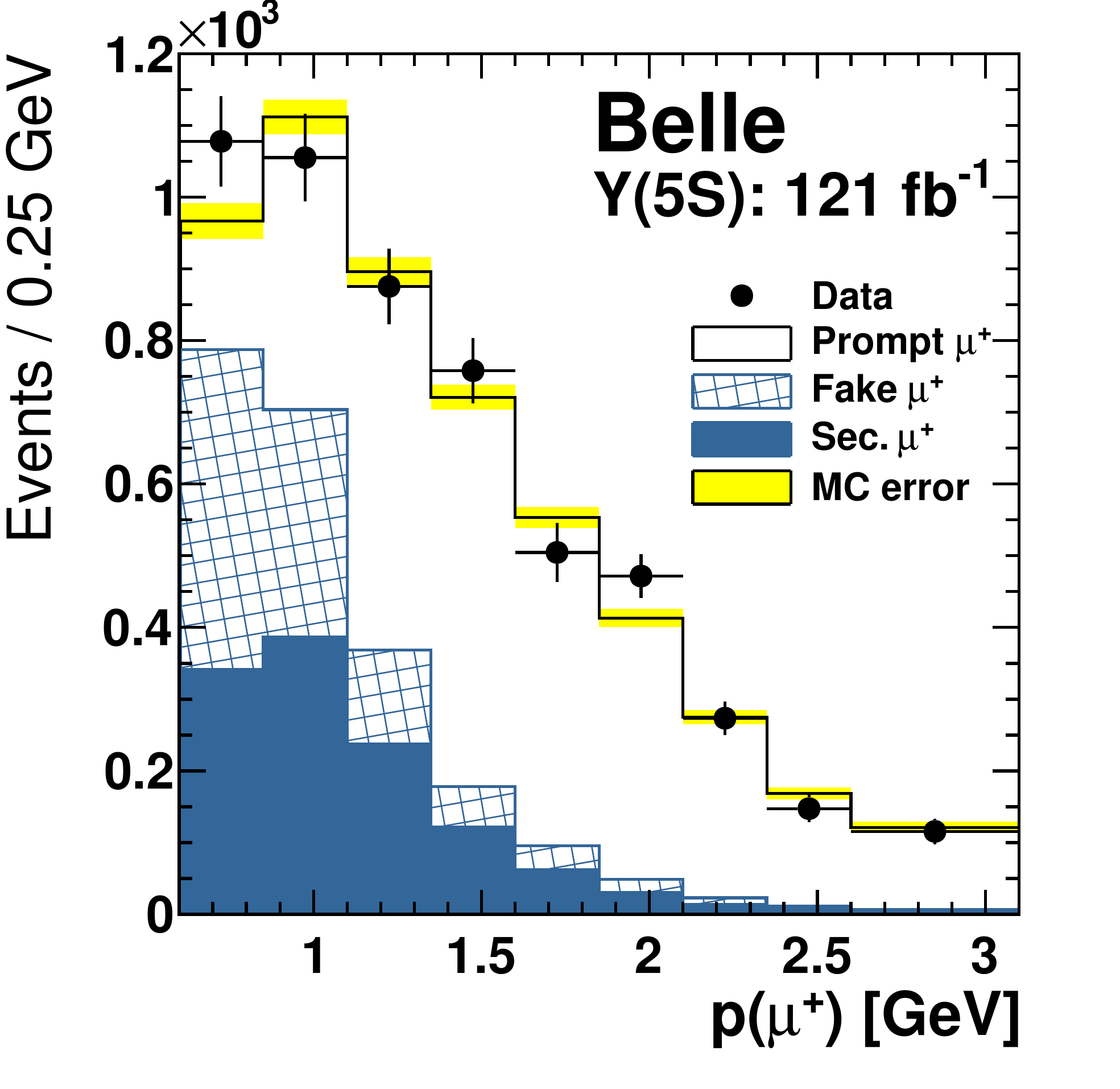}
\label{fig:muonmomfit}
}
\caption{
{\bf (a):} $KK\pi$ mass fits to the $D_s^+\ell^+$ samples collected near the $\Upsilon(5S)$ resonance. The figure shows $KK\pi$ mass for the whole range $p(\ell^+) > \unit[0.6]{GeV}$. {\bf (b) + (c):} Momentum spectra obtained from $KK\pi$ mass fits in bins of $p(e^+)$ and $p(\mu^+)$. Continuum backgrounds have been subtracted using off-resonance data. The MC uncertainty (yellow) comprises statistical and systematic uncertainties \cite{bs2xlnu}.}
\end{figure}

The Belle experiment collected a large data set of $\unit[121]{fb^{-1}}$ at a centre of mass energy of $\sqrt{s} = \unit[10.87]{GeV}$ near the $\Upsilon(5S)$ resonance. Only a fraction of $\Upsilon(5S)$ decays are into $B_s^{0(*)}\bar{B}_s^{0(*)}$ pairs ($f_s = (19.9 \pm 3.0) \%$), the rest is mainly into $B^+/B^0$ mesons. In order to enhance the relative abundance of $B_s^0$ decays in the selected events, we exploit the fact that $B_s^0$ decays proceed preferably via the Cabbibo favoured $\bar{B}_s \to D_s^+X$ transition with a multiplicity\footnote{The current PDG average is interpreted as multiplicity $\mathcal{B}(B_s^0 \to D_s^\pm X)$ and not as branching ratio $\mathcal{B}(B_s^0 \to D_s^- X)$.} of $\mathcal{B}(B_s^0 \to D_s^\pm X) = (93 \pm 25) \%$ \cite{PDBook}. $D_s^+$ mesons are reconstructed in the cleanest mode $D_s^+ \to \phi \pi^+,~\phi \to K^+K^-$. Signal lepton candidates $\ell^+$ ($\ell = e,~\mu$) are required to be of the same electric charge as the reconstructed $D_s^+$ meson to ensure that 
both stem from different $B_s^0$ mesons. The inclusive semileptonic branching fraction is extracted from ratios $N(D_s^+ \ell^+)/N(D_s^+)$, where $N(D_s^+)$ and $N(D_s^+\ell^+)$ are the efficiency corrected yields of $D_s^+$ mesons and $D_s^+\ell^+$ pairs in the sample determined from fits to the reconstructed mass of the $K^+K^-\pi^+$ system (see Fig. \ref{fig:kkpimassfit}). 

$D_s^+$ mesons from $c\bar{c}$ continuum background typically have high momenta and can be suppressed by the requirement $p^*(D_s^+)/p^*_{\text{max}}(D_s^+) < 0.5$, where $p^*$ denotes  momentum in the centre-of-mass frame of the $e^+e^-$ beams. The remaining background from $c\bar{c}$ decays is subtracted using a $\unit[63]{fb^{-1}}$ sample collected at a centre of mass energy of $\sqrt{s} = \unit[10.52]{GeV}$ below the energy threshold for the production of $B_{(s)}$ mesons (off-resonance). The background from secondary leptons, not coming directly from $B_{(s)}$ decays, and from misidentified leptons is estimated in a fit of the signal and background shapes derived from MC simulation to the lepton momentum distribution (see Figs. \ref{fig:elecmomfit} and \ref{fig:muonmomfit}). 

The measured ratio $N(D_s^+ \ell^+)/N(D_s^+)$ contains yields $\mathcal{N}_s$ from $B_s^0$ and $\mathcal{N}_{ud}$ from $B$ decays:
\begin{equation}
\frac{N(D_s^+ \ell^+)}{N(D_s^+)} = 
\frac{\mathcal{N}_s(D_s^+ \ell^+)+\mathcal{N}_{u,d}(D_s^+ \ell^+)}{\mathcal{N}_s(D_s^+) + \mathcal{N}_{u,d}(D_s^+)}~~(\ell=e, \mu)\,.
\label{eq:masterinc}
\end{equation}
The right side of the equation can be calculated from known production rates and branching fractions, and $\mathcal{B}(B_s^0 \to X \ell \nu)$. The branching fraction is extracted by simple algebraic  transformation of the equation. The external systematic uncertainties on the branching fraction are mainly due to the parameter $f_s/f_{u,d}$ describing the $\Upsilon(5S)$ hadronisation (3.2 \%) and branching ratios of the decays $B_s^0 \to D_s$ (4.4 \%) and $B \to D_s$ (3.2 \%). 

The results listed in Tab. \ref{tab:bsxlnu} are compatible with the previous measurement \cite{babarbsxlnu2012} and far more precise owing to a significantly larger dataset. Using the semileptonic width of the $B^0$ meson and the well measured $B_s$ and $B^0$ life times \cite{PDBook} the obtained semileptonic $B_s$ width $\Gamma_\text{sl}(B_s^0) = (1.04 \pm 0.09) \cdot \Gamma_\text{sl}(B_d^0)$ is in good agreement with theory predictions \cite{bigi2011,Gronau2010}.

\begin{table}
\centering
\begin{tabular}{lll}
\hline \hline
Mode & $N(D_s^+\ell^+) / N(D_s^+)~[10^{-4}]$ & $\mathcal{B}(B_s^0 \to X^- \ell^+ \nu_\ell)~[\%]$ \\
\hline
$e^+$ & $426 \pm 20_\text{stat.} \pm 13_\text{syst.}$ & $10.04 \pm 0.57_\text{stat.}\pm 0.37_\text{syst.} \pm 0.61_\text{ext.}$ \\
$\mu^+$ & $471 \pm 24_\text{stat.} \pm 16_\text{syst.}$ & $11.32 \pm 0.68_\text{stat.}\pm 0.46_\text{syst.} \pm 0.74_\text{ext.}$ \\
Combined & $446 \pm 16_\text{stat.} \pm 13_\text{syst.}$ & $10.61 \pm 0.46_\text{stat.}\pm 0.37_\text{syst.} \pm 0.67_\text{ext.}$ \\
\hline \hline
\end{tabular}
\caption{Preliminary results of the measured ratios and the extracted branching fractions. The combination of the electron and muon mode accounts for the correlations between the measurements.}
\label{tab:bsxlnu}
\end{table}

\section{Conclusion}
The branching fractions $\mathcal{B}(B^+ \to D_s^{(*)} K \ell \nu_\ell)$ were measured for the first time separately in the $D_sK$  and $D_s^{*}K$ modes. The measurements of the semi-inclusive decay modes are the first of their kind and shed new light on the ``inclusive vs. exclusive puzzle''. An important application of these measurements is in $B$ production studies, where they provide a precise normalisation. The inclusive semileptonic branching of the $B_s^0$ meson was measured to be $(10.6 \pm 0.5_\text{stat.} \pm 0.4_\text{syst.} \pm 0.7_\text{ext.})\%$ with a significant improvement in precision compared to the previous measurement \cite{babarbsxlnu2012}. 
 
\bibliographystyle{JHEP}
\bibliography{bibliography} 

\end{document}